\documentclass[twocolumn,trackchanges]{aastex63}
\usepackage{xcolor}

\usepackage[utf8]{inputenc}
\usepackage{textcomp}
\usepackage{amsmath,amssymb,gensymb,times,graphicx,morefloats,color, subfigure}
\usepackage{afterpage, bm}
\usepackage{natbib,hyperref}
\usepackage[outdir=./]{epstopdf}
\usepackage{mathrsfs}
\usepackage[normalem]{ulem}

\shorttitle{LTT 9779b secondary eclipses}
\shortauthors{Dragomir et al.}

\newcommand{\tess}{{\it TESS}}
\newcommand{\jwst}{{\it JWST}}
\newcommand{\kepler}{{\it Kepler}}

\newcommand{\spitzer}{{\it Spitzer}}
\newcommand{\lt}{{LTT9779}}

\newcommand{\ktwo}{{K2}}
\newcommand{\hst}{{\it HST}}

\begin{document}

\title{Spitzer Reveals Evidence of Molecular Absorption in the Atmosphere of the Hot Neptune \lt b}
\correspondingauthor{Diana Dragomir}
\email{dragomir@unm.edu}

\author[0000-0003-2313-467X]{Diana Dragomir}
\affiliation{Department of Physics and Astronomy, University of New Mexico, 1919 Lomas Blvd NE, Albuquerque, NM 87131, USA}

\author{Ian J.\ M.\ Crossfield}
\affiliation{Department of Physics and Astronomy, University of Kansas, Lawrence, KS, USA}
  
\author{Bj\"orn Benneke}
\affiliation{ Institute for Research on Exoplanets, Universit\'e de Montr\'eal, Montr\'eal, Quebec, H3T 1J4, Canada}  

\author[0000-0001-9665-8429]{Ian~Wong}
\affiliation{Department of Earth, Atmospheric and Planetary Sciences, Massachusetts Institute of Technology, Cambridge, MA 02139, USA}
\affiliation{51 Pegasi b Fellow}

\author[0000-0002-6939-9211]{Tansu Daylan}
\affiliation{Department of Physics, and Kavli Institute for Astrophysics and Space Research, Massachusetts Institute of Technology, Cambridge, MA 02139, USA}
\affiliation{Kavli Fellow}

\author[0000-0002-2100-3257]{Matias Diaz}
\affiliation{Departamento de Astronomia, Universidad de Chile, Camino El Observatorio 1515, Las Condes, Santiago, Chile}

\author{Drake Deming}
\affiliation{Department of Astronomy, University of Maryland, College Park, MD 20742, USA}

\author[0000-0003-4096-7067]{Paul Molliere}
\affiliation{Max-Planck-Institut f\"ur Astronomie, K\"onigstuhl 17, 69117 Heidelberg, Germany}
  
\author[0000-0003-0514-1147]{Laura Kreidberg}
\affiliation{Max-Planck-Institut f\"ur Astronomie, K\"onigstuhl 17, 69117 Heidelberg, Germany}
\affiliation{Center for Astrophysics $|$ Harvard \& Smithsonian, 60 Garden Street, Cambridge, MA, 02138, USA}

\author{James S. Jenkins}
\affiliation{Departamento de Astronomia, Universidad de Chile, Camino del Observatorio 1515, Las Condes, Santiago, Chile}
\affiliation{Centro de Astrofisica y Tecnologias Afine (CATA), Casilla 36-D, Santiago, Chile}

\author{David Berardo}
\affiliation{Department of Physics, and Kavli Institute for Astrophysics and Space Research, Massachusetts Institute of Technology, Cambridge, MA 02139, USA}

\author[0000-0002-8035-4778]{Jessie~L.~Christiansen}
\affiliation{Caltech/IPAC-NASA Exoplanet Science Institute, 1200 E. California Blvd. Pasadena, CA 91125}

\author[0000-0001-8189-0233]{Courtney~D.~Dressing}
\affiliation{Department of Astronomy, The University of California, Berkeley, CA 94720, USA}

\author{Varoujan Gorjian}
\affiliation{Jet Propulsion Laboratory, California Institute of Technology, Pasadena, CA, USA}

\author[0000-0002-7084-0529]{Stephen~R.~Kane}
\affiliation{Department of Earth and Planetary Sciences, University of California, Riverside, CA 92521, USA}

\author{Thomas Mikal-Evans}
\affiliation{Department of Physics, and Kavli Institute for Astrophysics and Space Research, Massachusetts Institute of
Technology, Cambridge, MA 02139, USA}

\author[0000-0001-9414-3851]{Farisa Y. Morales}
\affiliation{Jet Propulsion Laboratory,
California Institute of Technology, 4800 Oak Grove Drive, Pasadena, CA 91109, USA}

\author{Michael Werner}
\affiliation{Jet Propulsion Laboratory, California Institute of Technology, Pasadena, CA, USA}

\author[0000-0003-2058-6662]{George~R.~Ricker}
\affiliation{Department of Physics, and Kavli Institute for Astrophysics and Space Research, Massachusetts Institute of Technology, Cambridge, MA 02139, USA}

\author[0000-0001-6763-6562]{Roland~Vanderspek}
\affiliation{Department of Physics, and Kavli Institute for Astrophysics and Space Research, Massachusetts Institute of Technology, Cambridge, MA 02139, USA}

\author[0000-0002-6892-6948]{S.~Seager}
\affiliation{Department of Physics, and Kavli Institute for Astrophysics and Space Research, Massachusetts Institute of Technology, Cambridge, MA 02139, USA}
\affiliation{Department of Earth, Atmospheric and Planetary Sciences, Massachusetts Institute of Technology, Cambridge, MA 02139, USA}
\affiliation{Department of Aeronautics and Astronautics, MIT, 77 Massachusetts Avenue, Cambridge, MA 02139, USA}

\author[0000-0002-4265-047X]{Joshua~N.~Winn}
\affiliation{Department of Astrophysical Sciences, Princeton University, 4 Ivy Lane, Princeton, NJ 08544, USA}

\author[0000-0002-4715-9460]{Jon~M.~Jenkins}
\affiliation{NASA Ames Research Center, Moffett Field, CA, 94035, USA}

\author[0000-0001-8020-7121]{Knicole D. Col\'{o}n}
\affiliation{NASA Goddard Space Flight Center, Exoplanets and Stellar Astrophysics Laboratory (Code 667), Greenbelt, MD 20771, USA}

\author[0000-0003-0241-2757]{Willie Fong} 
\affiliation{Department of Physics, and Kavli Institute for Astrophysics and Space Research, Massachusetts Institute of Technology, Cambridge, MA 02139, USA}

\author{Natalia Guerrero}
\affiliation{Department of Physics, and Kavli Institute for Astrophysics and Space Research, Massachusetts Institute of Technology, Cambridge, MA 02139, USA}

\author{Katharine Hesse} 
\affiliation{Department of Astronomy, Wesleyan University, Middletown, CT 06459, USA}

\author[0000-0002-4047-4724]{Hugh P. Osborn} 
\affiliation{Department of Physics, and Kavli Institute for Astrophysics and Space Research, Massachusetts Institute of Technology, Cambridge, MA 02139, USA} 
\affiliation{NCCR/PlanetS, Centre for Space \& Habitability, University of Bern, Bern, Switzerland}

\author[0000-0003-4724-745X]{Mark E. Rose}	
\affiliation{NASA Ames Research Center} 		
\author[0000-0002-6148-7903]{Jeffrey C. Smith}	
\affiliation{SETI Institute/NASA Ames Research Center}

\author[0000-0002-8219-9505]{Eric B. Ting}	
\affiliation{NASA Ames Research Center}

\begin{abstract}

Non-rocky sub-jovian exoplanets in high irradiation environments are rare. \lt b, also known as TESS Object of Interest (TOI) 193.01, is one of the few such planets discovered to date, and the first example of an ultra-hot Neptune. The planet's bulk density indicates that it has a substantial atmosphere, so to investigate its atmospheric composition and shed further light on its origin, we obtained \spitzer\ IRAC secondary eclipse observations of \lt b at 3.6 and 4.5 $\mu$m. We combined the \spitzer\ observations with a measurement of the secondary eclipse in the \tess\ bandpass. The resulting secondary eclipse spectrum strongly prefers a model that includes CO absorption over a blackbody spectrum, incidentally making \lt b the first \tess\ exoplanet (and the first ultra-hot Neptune) with evidence of a spectral feature in its atmosphere. We did not find evidence of a thermal inversion, at odds with expectations based on the atmospheres of similarly-irradiated hot Jupiters. We also report a nominal dayside brightness temperature of 2305 $\pm$ 141 K (based on the 3.6 $\mu$m secondary eclipse measurement), and we constrained the planet's orbital eccentricity to $e < 0.01$ at the 99.7 \% confidence level. Together with our analysis of \lt b's thermal phase curves reported in a companion paper, our results set the stage for similar investigations of a larger sample of exoplanets discovered in the hot Neptune desert, investigations which are key to uncovering the origin of this population.

\end{abstract}

\section{Introduction}

To a large extent, the field of astronomy is the study of emergent radiation from distant objects in order to ascertain their compositions and physical properties.  When it comes to extrasolar planets, although thousands of planets are known to transit their host stars we often obtain the greatest insights into the much smaller set of planets whose thermal emission properties have been precisely measured.

Although the most precise thermal emission measurements come from directly imaged planets, these weakly-irradiated, long-period objects are structurally different than, and so only imperfect analogues for,  the atmospheres of highly-irradiated, short-period planets.  Thermal emission via secondary eclipse observations has been measured in dozens of hot Jupiters with few-band \spitzer\ \citep{Wer04} photometry. Though the wavelength coverage of such data is necessarily sparse, several trends have emerged. For example, hot Jupiters with $T_\mathrm{eq}\gtrsim2200$~K are consistent with zero Bond albedo and inefficient global heat redistribution, indicating bright, hot daysides and relatively cold, dark night sides \citep{Gar20}. Furthermore, hot Jupiters with $T_\mathrm{eq}\gtrsim1900$~K have consistent dayside brightness temperatures in the 3.6\micron\ and 4.5\micron\ \spitzer\ IRAC bandpasses \citep{Gar20, Bax20}.  
In addition, the hottest planets, such as the so-called ``ultra-hot Jupiters,'' also show qualitatively different atmospheres in which opacity from H$^-$, hydrides, and other non-oxides begin to play a much larger role than in the (relatively) cooler population \citep{Arc18,Lot18b,Eva19}.

Despite being relatively easy to discover in early surveys, the intrinsic occurrence of hot Jupiters is much lower than that of smaller planets \citep{How10a,Ful18b}. An apparently similar shortage of smaller, hot planets, deemed the hot Neptune desert, has been recognized more recently \citep{Sza11,Maz16}.  This desert may have formed through mass loss of sub-Jovian-sized planets, via mechanisms such as photoevaporation \citep{Owe18} or Roche Lobe Overflow \citep{Val15}.

\begin{deluxetable*}{c c c c c c c l}[t]
\hspace{-1in}\tabletypesize{\scriptsize}
\begin{center}
\tablecaption{\spitzer\ Observations of \lt \label{tab:obs}}
\end{center}
\tablewidth{0pt}
\tablehead{
\colhead{Channel} & \colhead{Integration time (s)} & \colhead{AOR Start [UT]} & \colhead{AOR End [UT]}  & \colhead{AOR No.} & \colhead{Aperture radius} & \colhead{$\beta$} & \colhead{Notes} \\
& & & & & \colhead{(pixels)} & }
\startdata
3.6 & 0.4 & 02/26/19 23:30 & 02/27/19 02:42 & 68701696 & 3.0 &  1.21 & Ch1 eclipse 1 \\
3.6 & 0.4 & 03/02/19 03:42 & 03/02/19 06:54 & 68697600 & 3.0 &  1.36 & Ch1 eclipse 2 \\
3.6 & 0.4 & 03/07/19 16:31 & 03/07/19 19:42 & 68694528 & 3.0 &  1.13 & Ch1 eclipse 3 \\
3.6 & 0.4 & 03/12/19 10:44 & 03/12/19 13:56 & 68699648 & 3.0 &  1.34 & Ch1 eclipse 4 \\
4.5 & 2.0 & 03/19/19 13:43 & 03/19/19 16:53 & 68695808 & 4.0 &  1.15 & Ch2 eclipse 1 \\
4.5 & 2.0 & 03/21/19 03:46 & 03/21/19 06:56 & 68700928 & 4.0 &  1.18 & Ch2 eclipse 2 \\
4.5 & 2.0 & 03/26/19 16:36 & 03/26/19 19:46 & 68697088 & 4.0 &  1.22 & Ch2 eclipse 3 \\
4.5 & 2.0 & 04/01/19 05:56 & 04/01/19 09:06 & 68695552 & 4.0 &  1.20 & Ch2 eclipse 4 \\
3.6 & 0.4 & 10/24/19 03:38 & 10/24/19 14:31 & 70006528 & 3.0 &  1.19 & Ch1 eclipse 5 \& $\frac{1}{2}$ phase curve\\
3.6 & 0.4 & 10/24/19 14:35 & 10/25/19 02:09 & 70007040 & 3.0 &  1.22 & Ch1, eclipse 6 \& $\frac{1}{2}$ phase curve\\
4.5 & 2.0 & 10/26/19 13:03 & 10/26/19 23:55 & 70005504 & 4.0 &  1.15 & Ch2, eclipse 5 \& $\frac{1}{2}$ phase curve\\
4.5 & 2.0 & 10/27/19 12:00 & 10/27/19 11:35 & 70006016 & 4.0 &  1.26 & Ch2, eclipse 6 \& $\frac{1}{2}$ phase curve\\
\enddata
\end{deluxetable*}

Because of the hot Neptune desert, few sub-Jovian planets with hot-Jupiter levels of irradiation ($T_\mathrm{eq}\gtrsim 1000$~K) are suitable for thermal emission measurements, and so such studies have largely focused on two disjoint sets: large, highly-irradiated hot Jupiters and smaller, cooler warm Neptunes \citep[e.g.][]{Wal19}. 

A new target excellently suited to bridge the gap between these two populations via secondary eclipse measurements of atmospheric composition and structure is the new ultra-hot Neptune \lt b (Jenkins et al. 2020). The \tess\ mission revealed this 4.6$\pm0.2$\,$R_\oplus$, 29.3$\pm0.8$\,$M_\oplus$ exoplanet on a 0.8~d orbit around its G dwarf host star, giving it an equilibrium temperature of roughly 2000~K.  The combination of intense irradiation and sub-Jovian size and mass (which imply a 9\% atmosphere by mass; Jenkins et al. 2020) makes \lt b a rare inhabitant of the hot Neptune desert, while its bright host star ($K_s=8.0$~mag) makes it the highest-S/N target for secondary eclipse measurements among known sub-Jovian planets. Measurements of the atmospheric composition of this desert-dwelling planet could shed light on the process by which it came to be, and on how this region of parameter space is cleared.  

\section{\spitzer\ Observations}

Observations of \lt\ were obtained in both {\it Warm} \spitzer\ IRAC channels \citep{Faz04}. Assuming a circular orbit for \lt b, we determined the most likely secondary eclipse times. For each channel, we scheduled \spitzer\ Astronomical Observing Requests (AORs) to overlap with four separate eclipse times \citep[GO: 14084;][]{crossfield:2018spitzer}. Two additional secondary eclipse time series (per channel) were obtained from AORs that covered a full phase curve of the planet in each channel \citep[which included two consecutive secondary eclipses; DDT 14290,][]{crossfield:2019ddt}. The dates of all the \spitzer\ observations are listed in Table~\ref{tab:obs}. The eight GO 14084 science AORs are 3.2 hours long. Each of the two GO 14290 phase curve time series spanned 22.5 hours. Exposure times were 0.4 s and 2 s for channel 1 and 2 observations, respectively.

We obtained the observations in subarray mode to reduce the data volume generated and better resolve short-timescale pointing jitter. Each eclipse-only science AOR was preceded by a 30-minute Pointing Calibration and Reference Sensor (PCRS) peak-up observation \citep{Ing12}. Since each of the two phase curve observations consisted of two {\it consecutive} science AORs, just one PCRS observation prior to the first AOR of each phase curve was sufficient. The PCRS peak-up AORs help minimize instrumental noise from intra-pixel sensitivity variations by eliminating the initial large drift that often occurs when the telescope slews to the target. The PCRS step also allows for the target to settle on the ``sweet spot" of a well-characterized pixel \citep{Ing12}. We do not use the 30-minute PCRS observations in our analysis.
Table~\ref{tab:obs} lists the details of all 12 of our Spitzer eclipse observations of \lt b.

\begin{figure*}[h]
    \centering
    \includegraphics[scale=0.48]{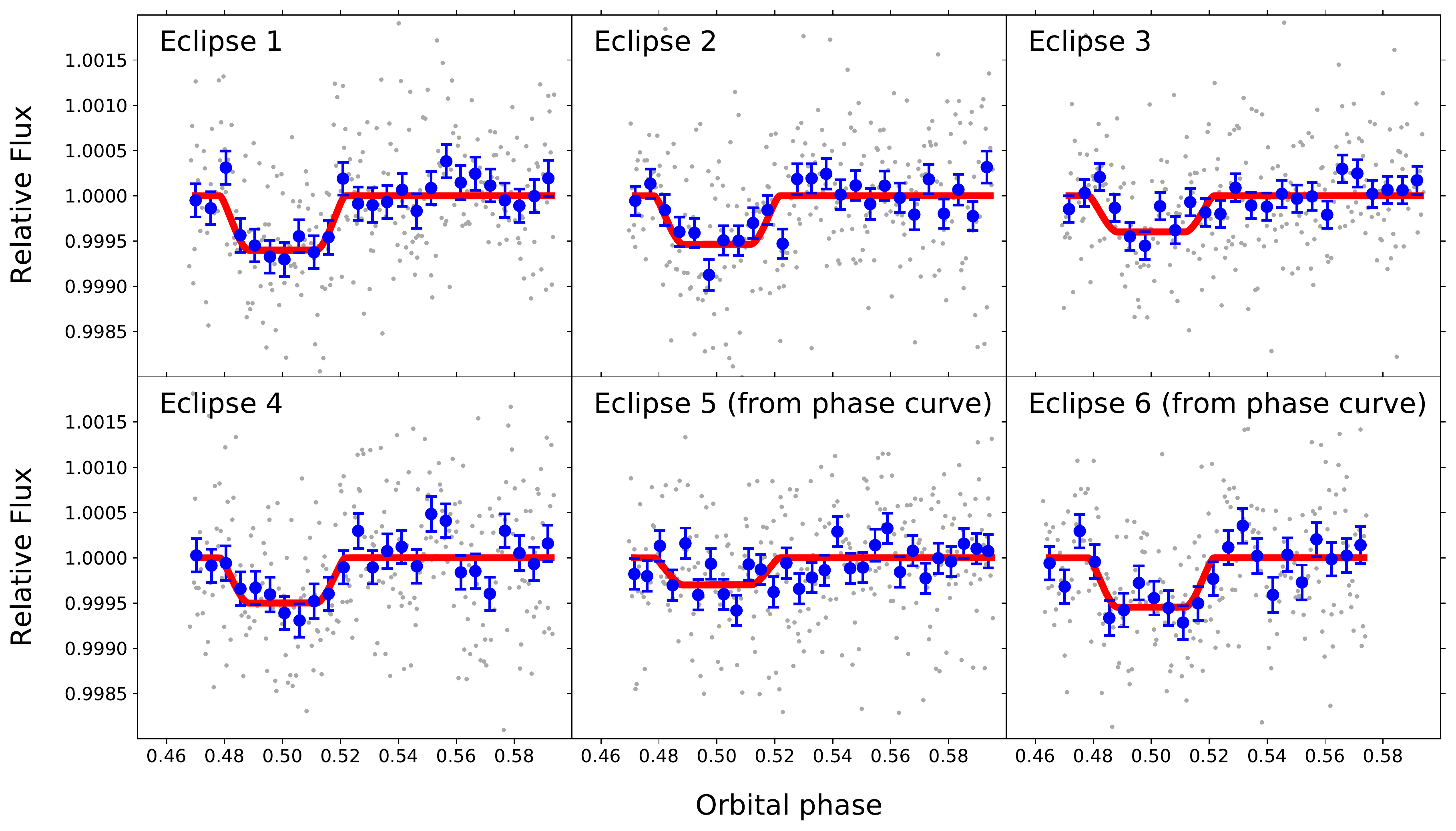}
    \caption{Individual \spitzer\ IRAC channel 1 (3.6\micron) eclipses of \lt b. Gray and blue points correspond to the unbinned and 6-min. binned corrected light curves, respectively. The best-fit model for each individual eclipse fit is shown in red.}
    \label{fig:indch1}
\end{figure*}

\begin{figure*}[h]
    \centering
    \includegraphics[scale=0.48]{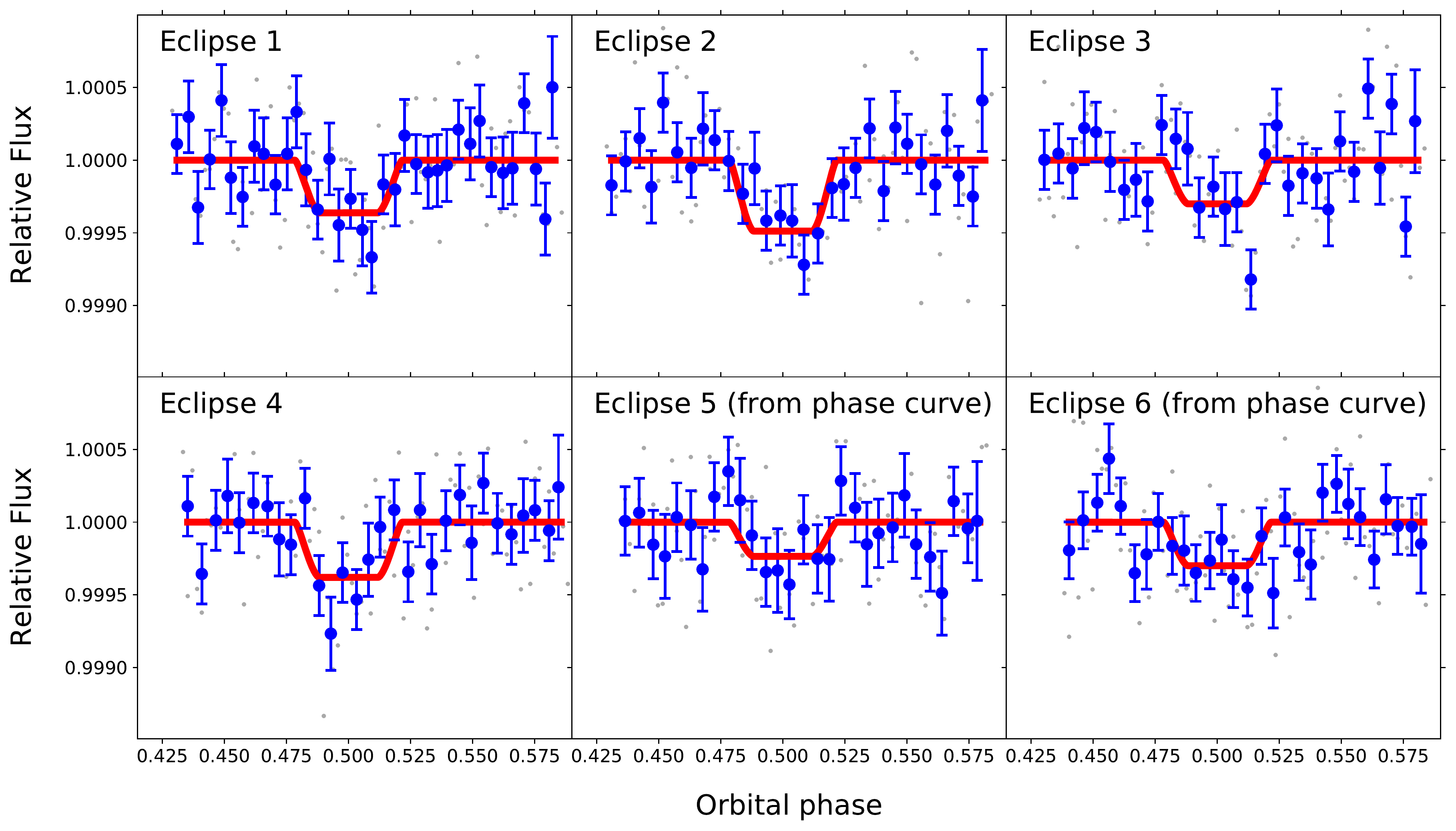}
    \caption{Individual \spitzer\ IRAC channel 2 (4.5\micron) eclipses of \lt b. Gray and blue points correspond to the unbinned and  6-min. binned corrected light curves, respectively. The best-fit model for each individual eclipse fit is shown in red. Note that the vertical scale differs from Fig. \ref{fig:indch1}.}
    \label{fig:indch2}
\end{figure*}

\begin{figure*}[t]
    \centering
    \includegraphics[scale=0.37]{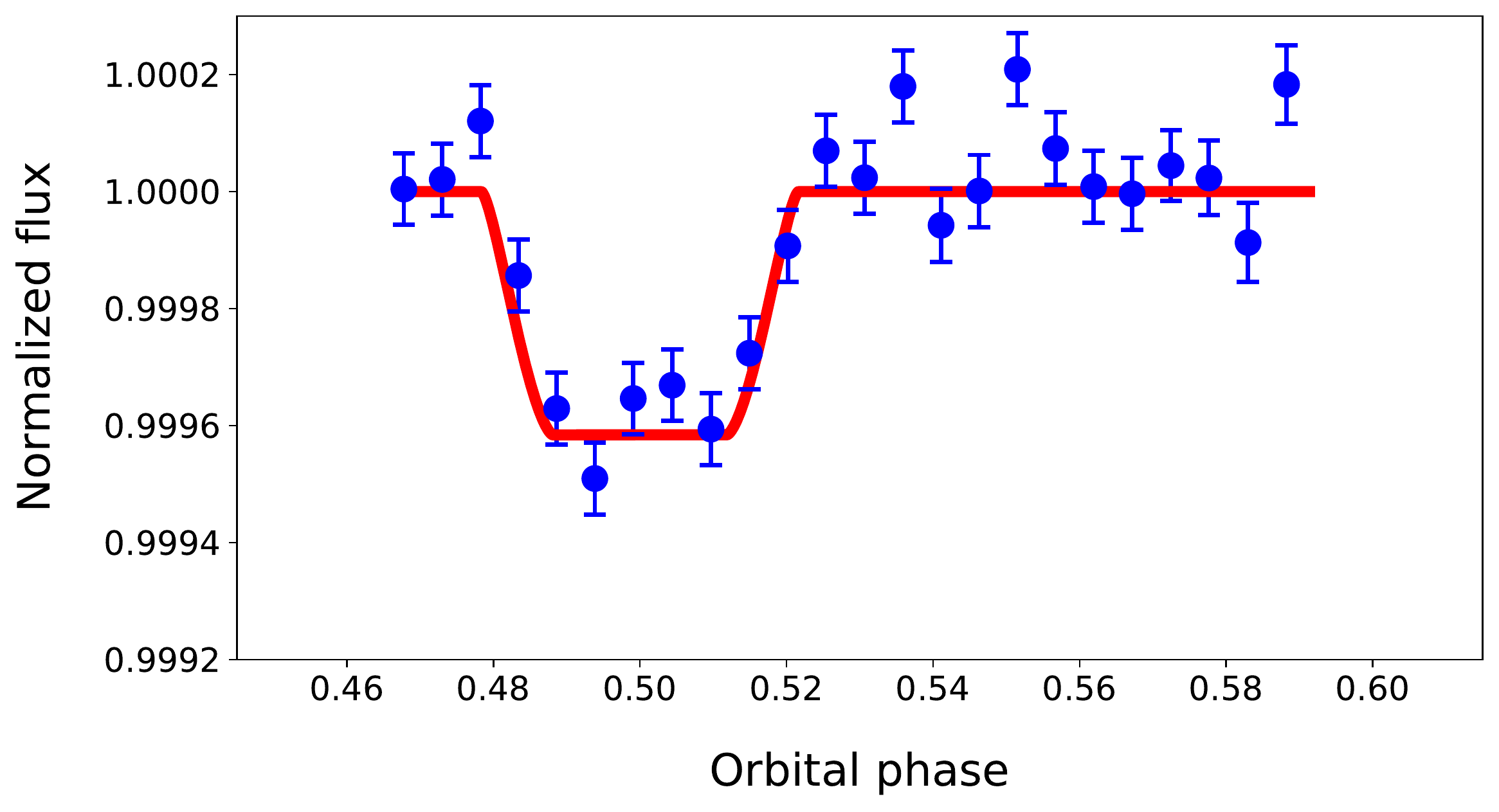}
    \includegraphics[scale=0.37]{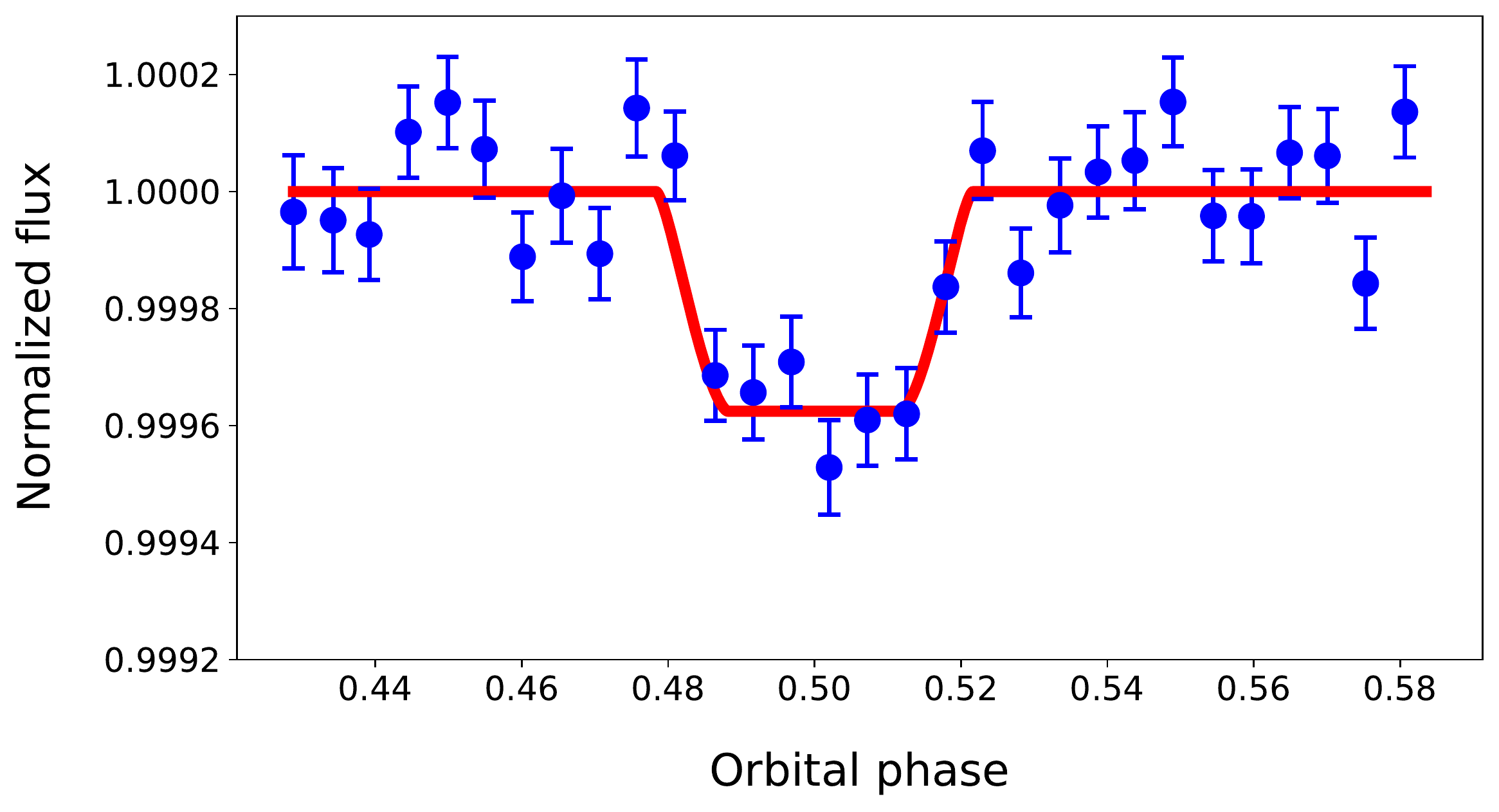}
    \caption{Phase-folded \spitzer\ IRAC channel 1 (left) and channel 2 (right) secondary eclipse  corrected light curves of \lt b. Blue points correspond to  6-min. bins. The best-fit model is shown in red.}
    \label{fig:jointch1ch2}
\end{figure*}

\subsection{Photometric Extraction and Light Curve Analysis}

We used the basic calibrated data (BCD) to extract light curves for a total of 12 eclipses, six in each channel. The BCD files are data cubes consisting of 64 images of 32 $\times$ 32 pixels. We follow a procedure similar to those described in previous works (e.g. \citealt{Ste12, Deming15}). For every image we subtract the sky background, we replace at each pixel position 4$\sigma$ outliers with the time median value for that pixel, and we use a 2D Gaussian fit to measure the centroid position of the stellar point spread function (PSF). Lastly, we sum the flux within eight circular apertures with a range of radii between 2 and 5 pixels.

For each 64-image data cube we reject 5$\sigma$ outliers from the cube median flux, and 10$\sigma$ outliers from the cube median x and y centroid position. We bin the remaining images in each data cube to obtain our raw light curve, and reject any points that are more than 4$\sigma$ from the median of the light curve flux. 

We began  the light curve analysis by trimming off the first 30 minutes of each one secondary eclipse light curve, which show a time-dependent ramp \citep{Demory11, Deming11} that is not well fitted by the Pixel Level Decorrelation (PLD) method \citep{Deming15} that we used to correct the systematics present in the light curves.

We used PLD to remove the correlated noise in the flux time series caused by pointing jitter combined with intra-pixel sensitivity variations of \spitzer\ IRAC. We select an array consisting of the 9 pixels that contain most of the flux from the star, and model the instrument systematics with a linear combination of the fluxes in those pixels (to correct the correlated noise due to the pointing jitter) and  either a first or second order polynomial in time (to correct the  component of the correlated noise that is due to temporal effects in detector sensitivity; \citealt{Demory12}):

\begin{deluxetable}{cccc}[h]
\tabletypesize{\footnotesize}
\tablecaption{\spitzer\ Secondary Eclipse Depths and Times \label{tab:spitzerdepths}}
\tablewidth{5cm}
\tablehead{\colhead{Eclipse ID} & \colhead{Wavelength} & \colhead{Eclipse depth} & \colhead{Mid-eclipse time (BJD)}}
\startdata
\vspace{2mm}
E1Ch1 & 3.6 & 600 $\pm$ 112 & 2458541.5374 $\pm$0.0011\\ 
\vspace{2mm}
E2Ch1 & 3.6 &  535 $\pm$ 128 & 2458544.7043$^{+0.0020}_{-0.0016}$\\
\vspace{2mm}
E3Ch1  & 3.6 & 450 $\pm$ 111 & 2458550.2491$^{+0.0023}_{-0.0026}$\\
\vspace{2mm}
E4Ch1 & 3.6 &  505 $\pm$ 125 & 2458555.0024 $\pm$ 0.0013\\
\vspace{2mm}
E5Ch1  & 3.6 & 320 $\pm$ 121 & 2458780.7396$^{+0.0051}_{-0.0067}$\\
\vspace{2mm}
E6Ch1 & 3.6 &  546 $\pm$ 123 & 2458781.5334 $\pm$ 0.0013\\
\textbf{Global Ch1} & \textbf{3.6} & \textbf{482 $\pm$ 47} & \textbf{2458541.53906}$\bm{^{+0.00083}_{-0.00074}}$ \\
\hline
\noalign{\vskip 2mm} 
\vspace{2mm}
E1Ch2 & 4.5 & 362 $\pm$ 131 & 2458562.1308$^{+0.0016}_{-0.0020}$\\ 
\vspace{2mm}
E2Ch2 & 4.5 & 480 $\pm$ 137 & 2458563.7159 $\pm$ 0.0017\\
\vspace{2mm}
E3Ch2 & 4.5 & 301 $\pm$ 176 & 2458569.2599$^{+0.0020}_{-0.0028}$\\
\vspace{2mm}
E4Ch2  & 4.5 & 348 $\pm$ 136 & 2458574.8022$^{+0.0023}_{-0.0029}$\\
\vspace{2mm}
E5Ch2 & 4.5 & 236 $\pm$ 184 & 2458783.1160$^{+0.0028}_{-0.0035}$\\
\vspace{2mm}
E6Ch2  & 4.5 & 282 $\pm$ 145 & 2458783.9049$^{+0.0034}_{-0.0040}$\\
\textbf{Global Ch2} & \textbf{4.5} & \textbf{372 $\pm$ 55} & \textbf{2458562.13302 $\pm$ 0.00092}
\enddata
\end{deluxetable}

\begin{equation}
\label{eqn:spitzersys}
\begin{split}
S(t) = \sum c_i P_i(t) + ft(+ gt^2),
\end{split}
\end{equation}
where $P_i(t)$ are the values of each of the 9 pixels 
as a function of time, $c_i$ are the coefficients associated with each $P_i(t)$ time series, $f$ (and $g$) are the linear (and quadratic) polynomial coefficients of the second order polynomial, and $t$ is the time. 
 We compute the Bayesian Information Criterion (BIC \citealt{Sch78}) for all individual eclipse fits, using either a first-order polynomial (BIC$_{1st}$) or a second-order polynomial (BIC$_{2nd}$). If $\Delta$BIC$=$BIC$_{2nd}-$BIC$_{1st}$ is positive, we determine that the former represents the data well; if $\Delta$BIC is negative, the latter is needed to adequately model the temporal detector sensitivity effects. A second-order polynomial is strongly preferred for the channel 1 light curves ($-41 < \Delta$ BIC $< -4$), while a first-order polynomial (i.e. linear function of time) is moderately preferred for the channel 2 light curves ($1 < \Delta$ BIC $< 4$). We model the light curves accordingly. While the planet's phase curve is present and detectable in the light curves (see Crossfield et al., in review for an analysis of the thermal phase curves), we find that employing a second order polynomial is sufficient to flatten the out-of-eclipse portions of the light curves without the need for computationally intensive phase curve fits.
In our systematics model, we omit the constant $h$ that \cite{Deming15} include in their systematics model because the first term of equation \ref{eqn:spitzersys} also serves to scale the out-of-transit level of the light curve, so this constant is not necessary \citep{Ben17}. We fit our data with the product of equation \ref{eqn:spitzersys} and the eclipse model ($E(t)$):

\begin{equation}
\label{eqn:spitzerall}
\begin{split}
F(t)= S(t) \cdot E(t) ,
\end{split}
\end{equation}

where $E(t)$ is analogous to the non-limb darkened transit model of \cite{Mandel}, except that the depth of the occultation represents the planet-to-star flux ratio (assuming the planet is a uniformly bright disk) rather than the planet-to-star area ratio. We fix the orbital period ($P$), scaled semi-major axis ($a/R_S$), orbital inclination ($i$) and planet-to-star radius ratio ($R_P/R_S$) to the values reported in Jenkins et al. (2020), which are constrained with much better precision by the 38 \tess\ transits they analyzed than we could achieve with the \spitzer\ eclipses. Specifically, we use $P = 0.7920520 \pm 0.0000093$, $a/R_S = 3.877^{+0.090}_{-0.091}$, $i = 76.39 \pm 0.43$ and $R_P/R_S = 0.0455^{+0.0022}_{-0.0017}$. Thus, we only fit for the mid-eclipse time and the eclipse depth $D$, as well as a scaling factor on the per-point photometric uncertainty $\beta$ (to ensure realistic uncertainties on the astrophysical parameters; e.g. \citealt{Pon06,Dem16b}). We use the Python package scipy.optimize with the L-BFGS-B method \citep{Byr95} to find a best fit solution which we use to initialize our MCMC algorithm, which uses the \texttt{emcee} code \citep{DFM13}. We perform the fit on light curves extracted from apertures with radii between 2 and 5 pixels. For our final results, we used the aperture that minimizes the photometric uncertainties and the correlated noise component (i.e. the fitted scaling factor mentioned above), which are those with 3 and 4 pixel radii, for channels 1 and 2, respectively. 

We fit each of the 12 eclipse light curves individually, and we also perform a global fit on the six light curves for each channel. Table \ref{tab:spitzerdepths} lists the best-fit values for the individual and global (in bold) eclipse depths and times from the MCMC runs. We also report the $\beta$ values in Table \ref{tab:obs}. Plots of the  systematics-corrected light curves are shown in Figures \ref{fig:indch1} and \ref{fig:indch2}. We show the combined light curves for channel 1 and 2 in Figure \ref{fig:jointch1ch2}.

Additional independent analyses were performed by three of our team members (two using their own versions of PLD and one using BLISS, \citealt{Ste12}), who found secondary eclipse values that agree with those shown in Table~\ref{tab:spitzerdepths} to within 0.6$\sigma$ (for each \spitzer\ channel). In particular, all analyses found shallower eclipses in channel 2.

\section{\tess\ Observations}

The \lt\ system was observed by the TESS spacecraft in Sector 2. To measure the TESS-band secondary eclipse depth and search for a possible visible-light phase curve signal from \lt b, we undertook two independent analyses of analyzed the Presearch Data Conditioning-Simple Aperture Photometry (PDC-SAP; \citealt{Smi12, Stu14}) light curve, which was processed through the Science Processing Operations Center (SPOC) pipeline \citep{Jen16} and has a 2-minute cadence. The PDC process is designed to correct the light curve for systematics while preserving astrophysical variability intrinsic to the target.

\subsection{Light Curve Analysis}

For the analysis we followed the methods described in detail in \citet{Shp19} and \citet{Won20}. We first removed all flagged data points from the time series and applied a 16-point-wide moving median filter to trim $3\sigma$ outliers. Then, the light curve was broken up into segments that are separated by the scheduled momentum dumps, which occurred ten times during Sector 2. We ignored the two short $<$0.5-day segments that immediately preceded the data downlink interruptions, arriving at a total of 10 segments. Segments 3 and 9 were affected by significant flux ramps prior to the subsequent momentum dumps, so we trimmed the last 0.75~day worth of data from those two segments. The full light curve used in our analysis includes 16433 points, and contains 29 transits and 30 secondary eclipses.

The system's normalized brightness variation was modeled as
\begin{equation}\label{full}
F(t)= S(t)\cdot \frac{1+\bar{f}_{p}-A\cos(2\pi\phi)}{1+\bar{f}_{p}},
\end{equation}
where $S(t)$ is a set of generalized polynomials in time used to model the long-term systematics trends (as defined in equation 5 of \citealt{Shp19})\footnote{When performing the brightness variation fits, we use polynomial orders that minimize the BIC.}; $\phi$ is the orbital phase (with the zeropoint set at mid-transit); and the variables $\bar{f}_{p}$ and $A$ represent the average relative brightness of the planet across its orbit and the semi-amplitude of the phase curve variation, respectively. By definition, the secondary eclipse depth is $D=\bar{f}_{p}+A$. We assumed zero eccentricity to fix the mid-eclipse time to $\phi$ = 0.5.

\begin{figure}
\begin{center}
\subfigure{\includegraphics[width=\linewidth]{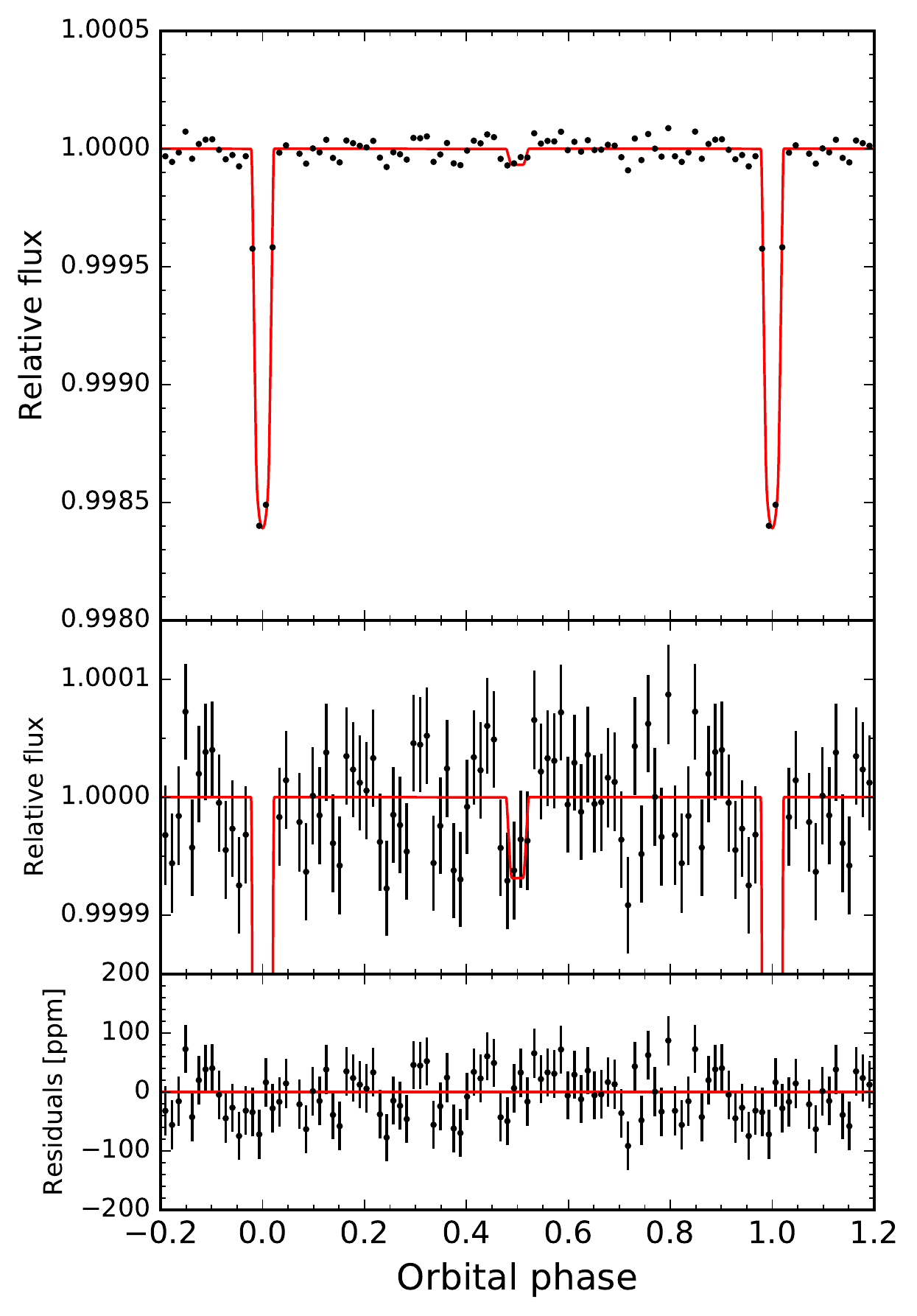}}
\caption{{\it Top panel:} \tess\ light curve of \lt\ after correction of the long-term systematics trends, phase-folded on the period of \lt b and binned in 15~minute intervals (black points). The best-fit full phase curve model from our joint analysis is shown by the red curve. {\it Upper middle panel:} zoomed-in view of the secondary eclipse and out-of-eclipse light curve. {\it Lower middle panel:} Corresponding residuals from the best-fit model.}
\end{center}
\label{fig:TESSfig1}
\end{figure}

We modeled the transits and secondary eclipses using \texttt{BATMAN} \citep{Kre15}, and the phase curve using Equation~\eqref{full}. We allowed both the orbital ephemeris ($T_{0}$, $P$) and the transit shape parameters ($b$, $a/R_{*}$) to vary freely. The median and $1\sigma$ uncertainties on all parameters were computed using \texttt{emcee} \citep{DFM13}.  The values of Jenkins et al. (2020) were used as initial guesses for the transit parameters, while $\bar{f}_{p}$ and $A$ were initialized at 100 and 0 ppm, respectively. To ensure that the reduced chi-squared value is near unity and produce realistic uncertainties on the astrophysical parameters, we included a scaling factor on the per-point photometric uncertainty ($\beta$) as a free parameter.

No significant phase curve signal was detected, with the best-fit amplitude $A$ being consistent with zero to within $1\sigma$. Therefore, for our final results we opted to fix the phase curve amplitude to zero and the time of secondary eclipse to phase 0.5. We measured a marginal secondary eclipse depth of $D=69_{-26}^{+28}$~ppm. The systematics-corrected, phase-folded and binned light curve is shown in the two upper panels of Figure~\ref{fig:TESSfig1}. 

We also performed an independent analysis of the \tess\ light curve using \texttt{allesfitter} \citep{Gue19b} following a methodology similar to that presented in \cite{Day19}, and found an eclipse depth of $D=59_{-21}^{+24}$~ppm. This is in very good agreement with the value reported above, which we thus use for the remainder of the paper.

\section{Atmosphere Modelling}

\begin{figure*}[bt]
\subfigure{\includegraphics[width = 0.5\textwidth]{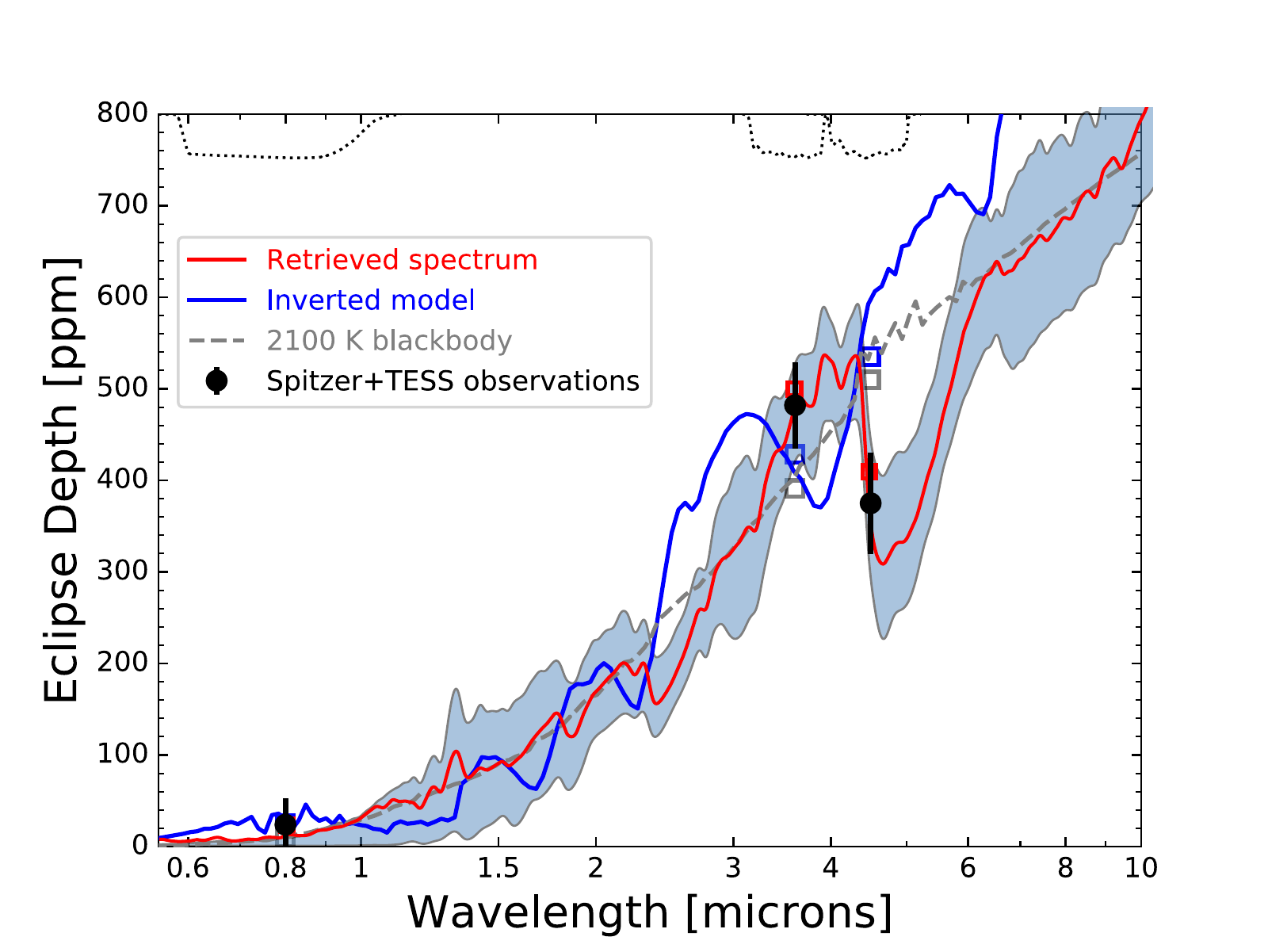}}
\subfigure{\includegraphics[width = 0.5\textwidth]{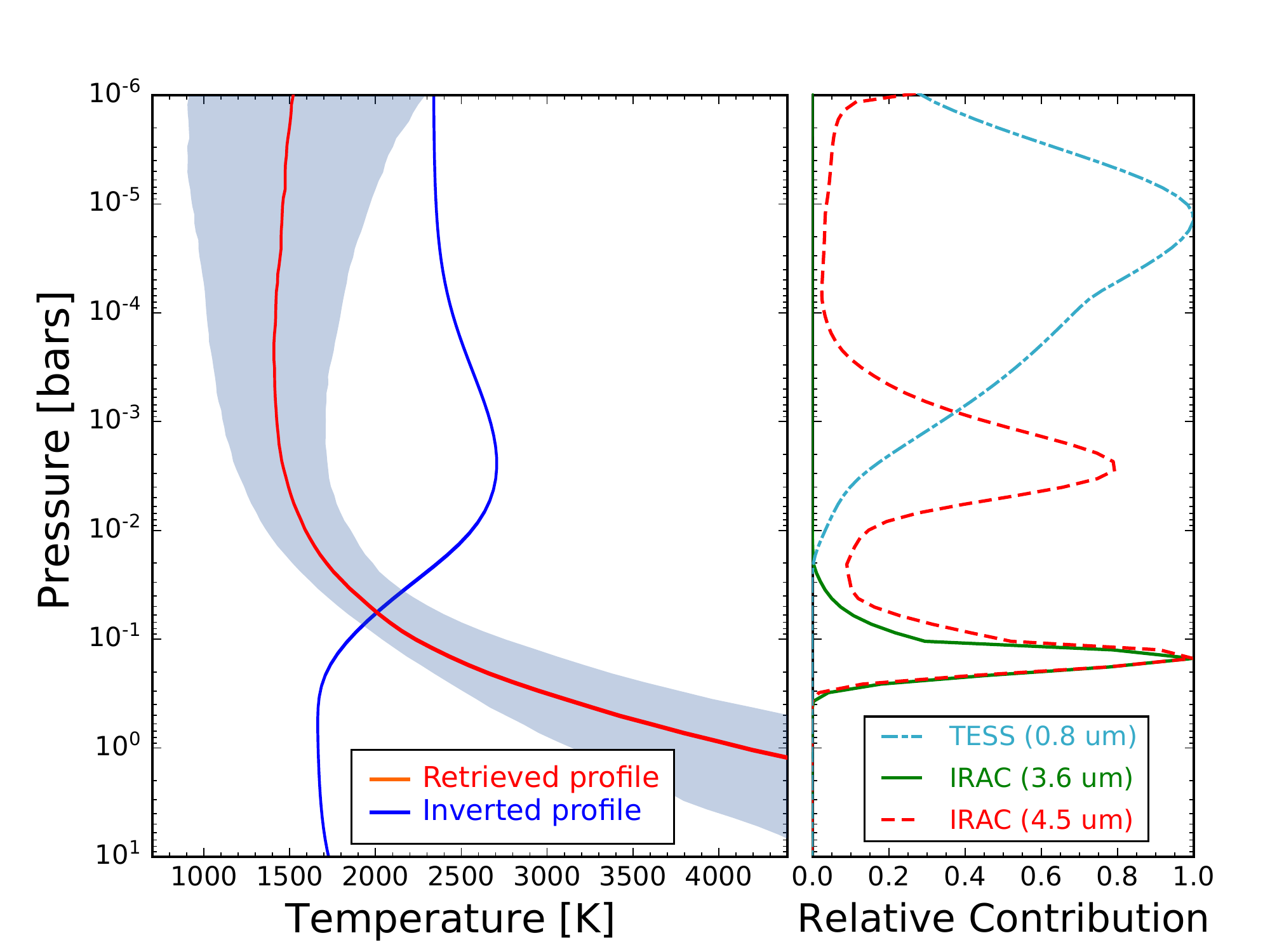}}
\caption{{\it Left:} Secondary eclipse measurements (black circles with error bars) with our weighted best-fitting model spectrum (red curve). The shaded region indicates the 68\% confidence interval from our retrieval analysis, and the dotted lines indicate the \tess\ and IRAC filter responses. The dashed line shows the planet/star flux ratios for a planet with blackbody emission at a temperatures of 2100~K. An inverted model is shown for reference (in blue). {\it Middle:} Retrieved thermal profile (solid line) with 68\% confidence interval (shaded region). The temperature decreases across the entire range of probed pressures. An inverted thermal profile is also shown (in blue). {\it Right:} Contribution functions for the \tess\ and IRAC bandpasses from the best-fit model spectrum. 
\label{fig:emissionspectrum}}
\end{figure*}

\subsection{Dayside brightness temperatures}

 Before analyzing the thermal emission spectrum of \lt b, we must take into account the fact that the planet's albedo likely contributes non-negligibly to the depth of the \tess\ secondary eclipse. We thus determine an upper limit on the thermal emission that can originate from the planet in the \tess\ bandpass. Since the 4.5 $\mu$m eclipse depth suggests molecular absorption at that wavelength (see below and section 4.2), we take the brightness temperature at 3.6 $\mu$m to be a more accurate estimate of the planet's dayside temperature. Using a stellar spectrum obtained with the BT-SETTL version of the PHOENIX atmospheric models \citep{All13}, we find a 3.6 $\mu$m brightness temperature of 2305 $\pm$ 141 K. For reference, the planet's equilibrium temperature is 1978 $\pm$ 19 K (Jenkins et al. 2020). Assuming a Bond albedo of 0 \footnote{While the \tess\ secondary eclipse depth suggests a geometric albedo $>$ 0 in the \tess\ bandpass, we cannot set constraints on the planet's overall Bond albedo with the available data. Thus, by using a Bond albedo of 0 we obtain an optimistic upper limit on the thermal emission in the \tess\ bandpass. \cite{Cro20} presents more detailed constraints on the planet's albedo and heat redistribution efficiency obtained from these brightness temperatures combined with observations of the planet's phase curve.} and that the planet re-emits all the flux it absorbs at the in the IRAC channel 1 bandpass, we find that the contribution from thermal emission to the \tess\ secondary eclipse is 27 ppm. We use this value for the analyses presented in the remainder of this Letter (but we note that even assuming a 0 ppm thermal emission contribution to the \tess\ secondary eclipse depth gives nearly identical results for the retrieval analysis we present in the next subsection).

We also fit a blackbody to the planet's three-point spectrum. We find the brightness temperature corresponding to the best-fit model to be 2100 $\pm$ 188 K (see dashed gray line in Figure \ref{fig:emissionspectrum}). However, the three eclipse depths are not well fitted by a blackbody emission spectrum. We interpret this as evidence of molecular absorption originating from the planet's atmosphere at 4.5 $\mu$m, which we investigate in detail below.

\subsection{Retrieval analysis} 

To interpret our \spitzer\ and \tess\ eclipses we used the free and open-source \texttt{petitRADTRANS} (pRT) radiative transfer and atmospheric modeling package \citep{Mol19}. We used the version of pRT available at its repository website\footnote{\url{https://gitlab.com/mauricemolli/petitRADTRANS/} as of April 2019}, which implements atmospheric retrieval. To speed up the retrieval algorithm, we adapted the online pRT code (which samples the atmospheric parameter space using Markov chain monte carlo techniques) to use nested sampling via the \texttt{MultiNest} algorithm \citep{Fer09,Buc14}. 

With just three data points, our retrieval cannot uniquely identify \lt b's atmospheric constituents, much less measure their precise abundances. However, we can rule out some combinations of models by virtue of their physical implausibility. For example, at the high temperature of this planet the CH$_4$  volume mixing ratio (VMR, used as a proxy for relative abundance) should be $\lesssim 10^{-4}$ and we expect a greater CO than CO$_2$ abundance, over a wide range of metallicity enhancements and C/O ratios \citep[e.g.,][]{Mos13,Hen16}. Thus, our retrieval parameters included as free parameters the (vertically-constant) atmospheric molecular mixing ratios for CO, H2O, TiO, VO, Na and K, as well as the parameters described in the analytical temperature-pressure profile of \cite{Gui10}. The best-fit models and 68\% confidence intervals are shown in Figure \ref{fig:emissionspectrum} for the emission spectrum (left) and the pressure-temperature profile (middle).

The highest volume mixing ratio we retrieve is for CO, at $-3^{+1.3}_{-1.7}$~dex, (which is driven by the deeper eclipse we observe at 4.5 $\mu$m relative to that observed at 3.6 $\mu$m). The H$_2$O VMR is poorly constrained since the two warm \spitzer\ IRAC channels alone are not very sensitive to this molecule (H$_2$O absorption is roughly equal in both channels). The VMRs of the optical opacity sources are similarly unconstrained\footnote{We find VMRs of -6.4$^{+1.9}_{-2.4}$, -6.3$^{+2.9}_{-2.5}$, -6.8$^{+3.0}_{-2.7}$, -5.5$^{+2.9}_{-3.0}$ and -5.7$^{+2.5}_{-2.9}$} dex for H$_2$O, TiO, VO, Na and K, respectively., likely due to degeneracies with the optical opacity parameter used in the \cite{Gui10} temperature-pressure profile. Our retrieval is therefore consistent with a wide range of atmospheric metallicities, but future measurements at higher precision and spectral resolution are needed to quantitatively constrain the atmospheric metal enhancement. Spectroscopy of the planet's atmosphere at shorter wavelengths can also better constrain the VMRs of the optical absorbers, if present.

To quantify the significance of the absorption feature at 4.5 $\mu$m, we calculate  BIC values for a 2100 K blackbody model and the model shown in red in Figure \ref{fig:emissionspectrum} (which includes CO absorption at 4.5 $mu$m). We find a $\Delta$BIC of 8, which indicates a strong preference for the CO absorption model. When including the \tess\ eclipse depth, we find a $\Delta$BIC of 12, which corresponds to a very strong preference for the model that include absorption at 4.5 $\mu$m.

The results of our retrieval also indicate that \lt b lacks a high-altitude thermal inversion. Our observations probe as deep as $\sim$1~bar (IRAC) to $\lesssim$1~mbar (\tess), and our retrieval shows no temperature increase across that pressure range (see right panel of Figure \ref{fig:emissionspectrum}). 
We note that we also performed a retrieval including CO$_2$, and found results in agreement with those presented above, except in this case there is a degeneracy between CO and CO2 abundances similar to those seen in other analyses of broadband hot Jupiter emission spectra \citep{Bar17,Wal19}. Similarly, assuming a 0 ppm thermal emission contribution to the \tess\ secondary eclipse depth (i.e. assuming that all of the \tess\ eclipse depth is due to reflected light, a physically allowed scenario since an albedo value of 1 would result in an optical eclipse depth of 137 ppm) results in retrieval values that differ only negligibly from those presented above (see also subsection 4.1). This is because the retrieval results are primarily driven by the (much more statistically significant) \spitzer\ eclipses.

\section{Orbital eccentricity}

We can use the timing of the \spitzer\ secondary eclipses to improve constraints on the planet's orbital eccentricity ($e$). On their own, they only constrain $e$cos$\omega$ (where $\omega$ is the argument of periastron), so we used \texttt{allesfitter} \citep{Gue19b} to perform a joint analysis of the radial velocity measurements (described in Jenkins et al. 2020), \tess\ transits, and \spitzer\ eclipses, which allowed us to fit both $\sqrt e$cos$\omega$ and $\sqrt e$sin$\omega$ (thus constraining $e$ and $\omega$ independently; \citealt{Alo18}). We place a 99.7\% (3$\sigma$) upper limit on $e$ of 0.01. 

\section{Discussion and Future Prospects}

While the means by which \lt b has retained its atmosphere remains a mystery (Jenkins et al. 2020), this very atmosphere now makes it the only sub-Jovian exoplanet with a detection of molecular absorption in its secondary eclipse spectrum to date. Of the most commonly expected molecules in hot exoplanet atmospheres, CO and CO$_2$ are the main absorbers at 4.5 $\mu$m (with H$_2$O a minor contributor). However, at the extreme temperatures of this planet, the former is likely to be significantly more abundant than the latter. We thus infer from our results the presence of CO in the atmosphere of \lt b. 

Most ultra-hot (T$_{eq} \gtrsim$ 2000 K) Jupiters with observed secondary eclipse spectra show a temperature inversion \citep[e.g.,][]{Chr10,Hay15,Eva17,She17,Kre18}, generally between 10 and 100 mbar, most often identified by a {\it positive} deviation of the 4.5 $\mu$m flux from a blackbody. \cite{Bax20} have empirically found that this transition between highly-irradiated gas giants with and without thermal inversions likely occurs at 1660 $\pm$ 100 K, well below the temperature of \lt b. So far only one is known to deviate from this trend (WASP-12b). To occur, these inversions generally require higher opacities in the optical than in the infrared, believed to be caused by the presence of optical absorbers such as TiO and VO \citep{Hub03}.  At these high temperatures, TiO and VO are believed to remain aloft at the low pressures probed at \spitzer\ IRAC wavelengths,  where they thus absorb a significant amount of flux and heat these upper layers of the atmosphere \citep{For08}. Our finding that \lt b lacks a temperature inversion thus further differentiates its atmosphere from that of most other hot Jupiters with similar irradiation levels. So why do we not observe a temperature inversion for this hot Neptune? Its temperature is not so high ($>$2500 K) so as to lead to the dissociation of TiO and VO \citep{Lot18b}. Instead, it could be that these optical absorbers are cold trapped on the planet's nightside \citep{Par13}, or that they exist in sub-solar abundances for other reasons. In the latter case, the lack of an inversion combined with the planet's high temperatures could potentially indicate that C/O $<$ 1, because a C/O around 1 favors the occurrence of inversions  even in the absence of absorbers like TiO and VO \citep{Mol15,Gan19}. Future detailed  Observations at higher spectral resolution (attainable with \jwst) and models including aerosols (which have been shown to impact the secondary eclipse spectra of hot Jupiter; \citealt{Par16, Tay20}) should be able to verify and refine the structure of the spectrum in this wavelength region, thus addressing at least some of these questions. 

The \spitzer\ observations alone cannot precisely constrain the atmospheric metallicity, but such a constraint could be achieved if they are combined with a measurement of water vapor absorption (within reach of \hst\ WFC3). Such a data set would also set the stage for determining the planet's atmospheric C/O. In the meantime, qualitative constraints on the metallicity may be inferred from the amplitude and phase offset of thermal planetary phase curves. Indeed, \spitzer\ phase curve observations of \lt b suggest a higher atmospheric metallicity than that of the typical hot Jupiter \citep{Cro20}. The planet's relatively high bulk density (1.677 $\pm$ 0.128 g cm$^{-3}$) also supports this hypothesis.

In the sub-Jovian regime, the atmospheric metallicities of the handful of sub-Jovian exoplanets for which this quantity has been measured show significant scatter spanning three orders of magnitude, even for masses $< 0.1 M_{Jup}$ \citep{Spake19}. Because of the small sample size, we cannot yet distinguish the relative roles of evolutionary history and birth environment in determining these planets' varied atmospheric compositions. \lt b can add a valuable new data point to this sample. Assuming it formed via mass-loss from an initially larger and more massive planet, its atmospheric metallicity could help determine to what extent mass-loss mechanisms such as photoevaporation preserve the metallicity of the primordial atmosphere. This is a compelling prospect, particularly since no quantitative theoretical predictions exist in the literature regarding the impact of mass-loss mechanisms on atmospheric metallicity. Spectroscopic measurements with higher precision and/or obtained over a wider range of wavelengths will improve the constraints on \lt b's atmospheric metallicity, probe the detailed composition of its atmosphere, and further investigate the absence of a thermal inversion.

Using the parameters from Jenkins et al. (2020) and equation 3 from \cite{Ada06}, we estimate the orbital circularization timescale for this planet to be 13-150 Myr, assuming a quality factor in the expected range (i.e. $10^5-10^6$; \citealt{Ada06}). Since the system is unlikely to be this young (Jenkins et al. 2020), we expect the orbit of \lt b to have been circularized by now, in line with our constraint on the planet's orbital eccentricity ($e < 0.01$ at 99.7\% confidence). This constraint does not point to the presence of an outer massive companion in the system that would excite \lt b's eccentricity, though a companion could still exist, as potentially indicated by the linear trend in the system's measured radial velocities noted by Jenkins et al. (2020). Continuing radial velocity monitoring and orbital obliquity measurements (e.g. via observations of the Rossiter-McLaughlin effect \citealt{Gaudi2}) could provide additional clues on the system's dynamical history and the presence of additional companions.

Given that even \kepler\ and \ktwo\ have not discovered other similar planets, \lt b may well remain peerless even after the end of the \tess\ survey. In any event, its unique location in the most isolated part of the hot Neptune desert makes it an invaluable target for comprehensive future characterization.

\section{Acknowledgments}

We are grateful to the referee for feedback that has improved the clarity of the paper and has prompted us to perform additional tests to verify our results. We thank James Owen for his thoughts on photoevaporation's impact on exoplanet atmospheres.

D. D. acknowledges support from NASA through Caltech/JPL grant RSA-1006130 and through the TESS Guest Investigator Program Grant 80NSSC19K1727. I. J. M. C. acknowledges support from the NSF through grant AST-1824644, and from NASA through Caltech/JPL grant RSA-1610091. TD acknowledges support from MIT's Kavli Institute as a Kavli postdoctoral fellow. M.R.D acknowledges the support of CONICYT/PFCHA-Doctorado Nacional 21140646, Chile. JNW thanks the Heising-Simons Foundation for support. C.D.D. acknowledges support from the Hellman Faculty Fund, the Alfred P. Sloan Foundation, and the David and Lucile Packard Foundation. JSJ acknowledges support through FONDECYT grant 1201371, and partial support from CONICYT project Basal AFB-170002.

This work is based [in part] on observations made with the \spitzer\ Space Telescope, which was operated by the Jet Propulsion Laboratory, California Institute of Technology under a contract with NASA. It is thanks to \spitzer's unique mid-IR capabilities combined with the overlap between its final year of operations and most of \tess' primary mission, that we have been able to obtain and present in this paper the first glimpse into a hot Neptune's atmosphere. 

Funding for the TESS mission is provided by NASA's Science Mission directorate. We acknowledge the use of public \tess\ Alert data from pipelines at the \tess\ Science Office and at the TESS Science Processing Operations Center. This research has made use of the Exoplanet Follow-up Observation Program website, which is operated by the California Institute of Technology, under contract with the National Aeronautics and Space Administration under the Exoplanet Exploration Program. Resources supporting this work were provided by the NASA High-End Computing (HEC) Program through the NASA Advanced Supercomputing (NAS) Division at Ames Research Center for the production of the SPOC data products.

\software{\texttt{allesfitter} \citep{Gue19b}, \texttt{ellc} \citep{Max16}, \texttt{dynesty} \citep{Spe20}, \texttt{emcee} \citep{DFM13}, \texttt{batman} \citep{Kre15}, \texttt{matplotlib} \citep{Hun07}, \texttt{numpy} \citep{Wal11}, \texttt{scipy} \citep{Vir20}.}

\facilities{\spitzer, \tess.}

\bibliographystyle{aasjournal}
\bibliography{research}

\end{document}